# Strain rate effects in the mechanical response of polymer anchored carbon nanotube foams

A. Misra[1], J.R. Greer[2], C. Daraio[1,3]

[1] *Graduate Aeronautical Laboratories (GALCIT),*

[2] *Materials Science*

[3] *Applied Physics*

*California Institute of Technology, Pasadena, CA, 91125*

**Super-compressible foam-like carbon nanotube films[1-7] have been reported to exhibit highly nonlinear viscoelastic behaviour in compression similar to soft tissue.[4] Their unique combination of light weight and exceptional electrical, thermal and mechanical properties have helped identify them as viable building blocks for more complex nanosystems and as stand-alone structures for a variety of different applications. In the as-grown state, their mechanical performance is limited by the weak adhesion between the tubes, controlled by the van der Waals forces, and the substrate allowing the forests to split easily and to have low resistance in shear.[5] Under axial compression loading carbon nanotubes have demonstrated bending, buckling[8] and fracture[9] (or a combination of the above) depending on the loading conditions and on the number of loading cycles[4]. In this work, we partially anchor[10] dense vertically aligned foam-like forests of carbon nanotubes on a thin, flexible polymer layer to provide structural stability, and report the mechanical response of such systems as a function of the strain rate. We test the sample under quasi-static indentation loading and under impact loading and report a variable nonlinear response and different elastic recovery with**

**varying strain rates. A Bauschinger-like effect is observed at very low strain rates while buckling and the formation of permanent defects in the tube structure is reported at very high strain rates. Using high-resolution transmission microscopy we report for the first time carbon nanotube mechanics where inner walls delaminate and crumble. These polymer-anchored CNT foams are reported to behave as conductive nanostructured layers, suitable as fundamental building blocks for a variety of different applications, or as new self-standing application-ready materials with potential employment as actuators, impact absorbers, or as layered components for the creation of acoustic metamaterials.**

Because of the excellent thermal, electronic and mechanical properties, vertically aligned carbon nanotube (CNT) arrays have been proposed for several potential applications, ranging from bio-mimetic adhesives similar to spider's and gecko's feet,[11] to nanobrushes,[12] vibration damping layers,[6] and multifunctional composites,[13] but their development into successful commercial applications has been limited by their weak adhesion to the growth substrate, resulting in poor resistance to shear. In the present work we grew long, vertically-aligned multiwall CNTs (Fig. 1a), transferred and anchored them in thin polymer layers (Fig. 1b), and tested their mechanical response. The goal of the anchoring was to create versatile and reusable nanosystems with tremendous flexibility, which integrates the excellent nanotubes' properties in a portable and structurally stable system. The arrays of long multiwalled carbon nanotubes were grown using a thermal CVD system on Si substrates (see Methods). Previous investigations of cyclic compressive loading of CNTs foams[3] reported that such structures have a slightly anisotropic mechanical response between the tips and the base of the tubes, with the base part being more prone to buckling (and therefore more inclined to demonstrating a nonlinear response) due to a lower overall density. Our anchoring method is designed to embed only the tips of the tubes into the polymer,[10] leaving the bases exposed to the indenter or a ball contact during mechanical testing and therefore maximising the observed nonlinear effects. A zoomed-in view of the thin polymer anchoring layer is provided in the lower inset of Fig. 1b.

Flat punch indentations (Fig. 2) and drop ball impact tests (Fig. 3) were performed to characterize their quasi-static and dynamic response in compression. Indentation measurements (Fig. 2a) were obtained using a flattened (by focus ion beam) Berkovich diamond punch. The tests were performed using DCM module of the MTS Nanoindenter G200, in continuous stiffness measurement (CSM) mode at room temperature, varying the displacement rate during loading (for details, see Methods). The load-displacement data curves are presented in Fig. 2b. It is evident from the curves that there are three distinct regions upon loading, most likely related to densification, bending and buckling modes of the nanotubes immediately under the indenter. This nonlinear behaviour is qualitatively similar to the viscoelastic properties reported for tests of single attached myoblasts cells under compression[14] and soft open foams.[15] We report that the amount of elastic recovery is inversely proportional to the indentation depth (varying between 10% and 25% in the tested range). The load-displacement curves were analyzed using the flat punch/infinite medium contact analysis method developed by Sneddon[16] with the projected area of compression being that of the nanoindenter flat punch. We obtained a value of the CNTs foams Young's modulus of ~0.4 MPa. Considering the forest density of ~100 CNTs/μm$^2$ and the area of applied load corresponding to ~30 μm-diameter flat punch cylinder, this value is much lower than the previously reported modulus of ~50 MPa[3] for a uniformly compressed freestanding forest of CNTs. Assuming a purely uniaxial compression we plotted normal stress-strain curves with varying loading/unloading cycles. It is very interesting to notice that the nanotube foam consistently exhibits a Bauschinger-like effect during the unloading-loading paths (Fig. 2c). To quantify the strain rate effects on the hysteretic response we plotted the size of the Bauschinger effect as a function of strain. It is evident that the hysteresis increases proportionally to both the pre-strain and the strain rate. We explain this phenomenon by the local densification effects directly below the compressed area. This finding is consistent with the previous reports of the viscoelastic compressive response and densification effects in free-standing non-anchored structures under uniform applied stress.[4] While the nanotubes appear to be vertically aligned throughout the entire thickness of the foam (Fig. 1b), SEM images taken at higher magnifications (see upper inset of Fig. 1b) reveal that the nanotubes are rather entangled in an open foam-type microstructure throughout the sample thickness. Such cellular structure at the microscale is responsible for the soft and compliant response observed in our tests, confirming the similarity of the anchored nanotube forests to the typical behaviour of foams.[15]

We investigated the structural anisotropy of the anchored nanotubes foams by monitoring the conductivity of the foams in the in-plane and the cross-sectional orientation (including the PDMS anchoring layer). The conductivity at the foam's surface was measured to be 0.42 $Scm^{-1}$ while that along the nanotubes length and across the thin polymer 0.16 $Scm^{-1}$, demonstrating that the CNTs go through the polymer layer leaving most of their tips exposed on the opposite side. Such a layer of polymer/CNTs composite, therefore, allows for conduction through the otherwise insulating polymer with a conductivity value only slightly lower than the previously reported value for an as-grown CNT forest.[4] The reported conductivity of the polymer-anchored nanotube foam opens the door for a myriad of applications ranging from nanoactuators to chemical separators membranes and sensing devices.[17]

To evaluate the high strain rate response of the polymer anchored foams, we performed drop-ball impact testing (see Methods) while systematically varying the impact velocity. Results related to the highest impact velocity (4 m/s) are reported in Fig. 3. Such impact velocity roughly corresponds, for example, to the drop of an electronic device (i.e. cell phone, remote or a personal computer) from an average height of a table/shelf. From the Force ($F$)-time ($t$) responses reported in Fig. 3b it is evident that the anchored nanotube forest works efficiently as an impact absorber and a pulse mitigation layer, suggesting its applicability as a free-standing protective layer in microelectronic packaging. To show their effectiveness we compare the impact mitigation performance of the polymer-anchored CNTs (curve 3) with the same impact performed on a single layer of polymer with no nanotubes (curve 1) and on an as grown CNTs forest on a Si substrate (curve 2). Note also that in the latter the nanotubes forest is flipped upside down (with tips headed up) with respect to the anchored layer reported in curve 3. The difference reported here is striking also when comparing the Force ($F$)-displacement ($\delta$) response under impact. The anisotropic response of the nanotube foams upon tip or base impact is evident (compare curves 2 and 3 in Fig. 3c), showing a more pronounced nonlinear response in the latter.

We evaluated the recovery and permanent deformation damage by using scanning and transmission electron microscopy (Fig. 4). The effect of the flat indentation tests on the surface of the foamlike forests of CNTs is shown in Fig. 4a. The inset highlights a cross-sectional view, etched with a focused ion beam, of the nanotubes foam below the indenter mark. The bending and the loss of orientation of the initially uncompressed open nanotubes foam are evident. The diameter of the circular indentation mark is ~30 μm, which matches exactly the area of the cylindrical indenter. The impact area (~1 mm in diameter) after a 4.0 m/s drop ball test is reported in Fig. 4b. In light of the large deformations reported in the force-displacement behaviour (Fig. 3c) it is evident that the nanotube foam is capable of a very large spring-back recovery (from a maximum compression of ~600 μm), leaving the surface of the film only partially damaged. The maximum local pressure in the impacted area has been calculated at ~60 MPa. Such high stresses are likely to cause locally permanent damage to the tubes, which we investigated via high resolution transmission electron microscopy (FEI TF30U). Figure 4c shows the microstructure of a typical undamaged, as grown carbon nanotube in the forest. The effects of the highest velocity impact on the buckled tubes are reported in Figs. 4d and 4e. We noticed two different types of permanent damage of the structure: in addition to the previously reported bending and rippling[18] of the tubes (Fig. 4e) we discover a new effect of delamination and crumbling of the inner walls of the tubes (Fig. 4d), probably related to the confining effect provided by the outer walls upon impact. To the best of the authors' knowledge this represents the first experimental report of such dynamically generated defect and challenges some of the classical theoretical and numerical prediction of carbon nanotubes deformation at high strain rates, opening up new avenues for computational investigations.

In conclusion, anchored foam-like forest of carbon nanotubes were found to demonstrate a highly nonlinear dynamic response when subjected to mechanical impact as well as excellent energy absorption capabilities. At small strain rates (on the order of $10^{-8}s^{-1}$) the response of the anchored foams appears to be elastic-plastic with the hysteretic loading/unloading response sensitive to the variation in the strain rate. At higher strain rates ($10^3$-$10^4$ $s^{-1}$) and axial loads, the formation of permanent defects in the multiwalled structure of the CNTs in the foam is reported and suggests new modes of deformation related to the delamination of the tubes' cores. These results suggest that foamlike forests of CNTs strongly anchored in thin polymer layer form hybrid structures between pure CNTs forests and CNTs composite films,[19] with significantly enhanced properties over their individual components, providing a viable engineering solution for light weight, small shock absorbers and impact protective layers for electronics and space applications.

## METHODS

### CNTs GROWTH AND ANCHORING

The arrays of multiwalled carbon nanotubes were grown on Si substrates using a two-stage thermal CVD system. The solution of catalyst (ferrocene) and carbon source (toluene) was heated at 825 ºC in a long quartz tube, in the presence of argon flow as carrier gas. The length of the grown forest was ~800 µm. A rapid transfer method from the growth substrate to the thin polymer layer has been employed. Polymer polydimethylsiloxane (PDMS) was spin-coated on top of the glass slide at 800 rpm to get a ~50 µm thick film. The carbon nanotube forests could then be anchored on top of the polymer surface. The polymer was cured after partial infiltration at 80 ºC temperature for 1 h. After, the anchored films were peeled off the glass slide. The advantage of this method is that the geometry of the nanotubes network can be predetermined by the growth conditions in the CVD chamber. Embedded carbon nanotubes showed excellent vertical alignment with an average height of ~750 µm.

### FLAT INDENTATION MECHANICAL TESTING

Indentations were performed using the dynamic contact module of a MTS Nanoindenter G200 with a flat punch indenter tip. The flat punch tip was custom fabricated from a standard Berkovich indenter by using the FIB to machine off the diamond tip, resulting in the projected area of a circle with a ~30 µm inscribed diameter. The MTS G200 Nanoindenter system is thermally buffered from its surroundings to within $1^{o}$C; however, small temperature fluctuations cause some of the machine components to expand and contract, and this thermal drift is corrected by monitoring the rate of displacement in the final 100 seconds of the hold period. Load-displacement data were collected in the continuous stiffness measurement (CSM) mode of the instrument. The experimental procedure involved first, locating the area of choice under the top-view 40X optical microscope, then calibrating the indenter to microscope distance to within a fraction of a micron on the surface of the sample away from the selected position, and finally moving the calibrated flat indenter tip to the position directly above the selection. Thermal drift stabilization follows the compression of the foam at a constant nominal displacement rate. During the initial segment of the test, the instrument locates the sample surface and then moves to the specified location and starts the initial approach segment, decreasing the approach velocity to 54nm/sec when the indenter is less than 2 μm above the surface. Once the surface of the CNT foam has been detected, such parameters as the load, or force, harmonic contact stiffness, and the compressive displacement of the surface from the point of contact are continuously measured and recorded.

## IMPACT TESTING

The experimental setup used for high strain rate tests consisted of a benchtop system[5,6,8] which included a free-falling sphere (Bearing-Quality Aircraft-Grade 25, Alloy Chrome Steel precision stainless steel ball, diameter 4.76 mm, with a surface roughness (RA) ~50 nm maximum, made from AISI type 52100 steel, McMaster-Carr cat.) and a calibrated piezosensor (Piezoelectric single sheet, T110-A4-602 provided by Piezo-System, Inc. with soldered 34 AWG microminiature wiring) connected to a Tektronix oscilloscope (TDS 2024B) to detect force-time curves to help control the overall dynamic force applied to the CNTs foams. The use of a sphere, as opposed to a flat plate, allows the application of a reproducible large concentrated force so that each nanotube can be subjected to sufficient impact energy. The impact on the aligned nanotubes was generated by dropping the 4.76 mm diameter steel sphere (0.45 g) from variable heights (0.5-80 cm), which corresponds to a speed of impact of ~0.3-4 m/s. Accordingly, the overall strain rate was calculated to be on the order of $10^3$-$10^4$ $s^{-1}$.

## POLYMER-CNTs ADHESION TESTING

We have performed independent tension tests on doubly-anchored CNTs forests (two PDMS layers were anchored on both the top and bottom of the forest) to evaluate the effective adhesion of the CNTs with the anchoring polymer layer. To ensure uniform gripping for the tests, the PDMS layers were first spin-coated on glass slides and then cured. We used a custom made tension/compression test system uses an ALD-MINI-UTC-M 500g load cell from A.L. Design. Tension test results showed consistently that the maximum normal tension force at failure was measured ~2.3 N and in all cases failure always happened by the detachment of the PDMS polymer from the glass slide, and never by debonding of the CNTs from the anchoring layer. These results are consistent with what reported for similarly anchored CNTs in RTV layers[10] and confirm the excellent adhesion of the tubes with the thin substrate.

## ELECTRICAL TESTING

Two-point electrical measurements were performed by using an Alessi, REL-3200 probe station, attached with Keithley-236 source measure unit system to evaluate the in plane conductivity at the surface of the CNTs array as well as along the tubes length and through the anchoring PDMS polymer layer. A constant current (5 mA) was applied while the voltage was measured.

**Acknowledgements** C.D. and J.R.G. wish to acknowledge the support of this work by their Caltech start-up funds, A.M. acknowledges support by the Moore Fellowship. The authors also thank C. Kovalchick for his support on the CNTs/polymer adhesion tests and C. Garland on TEM supervision.

**Competing interests statement** The authors declare that they have no competing financial interests.

**Correspondence** and requests for materials should be addressed to C.D. (e-mail: daraio@caltech.edu).

**Figure 1.** Synthesis and assembly of the transportable polymer anchored CNTs nanofoams. **a**, Schematic diagram showing the growth and anchoring steps for the CNTs forests. **b**, Scanning electron micrograph of the nanotubes films showing the alignment. Top inset shows a higher magnification image of the microstructure of the foam. Bottom inset is a zoom-in of the polymer anchoring layer. The nanotubes tips are embedded in the polymer going fully through the polymer thickness as confirmed by electrical measurements.

**Figure 2.** Flat punch nanoindentation results. **a**, Schematic diagram showing the experimental set up. **b**, Load-displacement curves obtained at different loading rates. **c**, Stress-strain curves extrapolated by the indentation measurement at varying strain rates upon various loading/unloading cycles showing the presence of a Baushinger-like effect. **d**, Dependence of the hysteresis loops amplitude on strain.

**Figure 3.** Impact (high strain rate) results. **a**, Schematic diagram showing the experimental set up. **b**, Load-displacement curves obtained impacting a stainless steel bead at ~4 m/s on a single PDMS layer (curve 1), an as grown CNTs forest on a silicon substrate (curve 2) and on our PDMS anchored CNTs forest (curve 3). **c**, Force-time response measured experimentally for the same impacts.

**Figure 4** Characterization of the deformed forests. **a**, SEM image obtained on the surface of the sample after flat indentation tests. The inset shows the cross sectional view of the area underneath the indenter after FIB slicing. **b**, SEM image of the damaged area on the carbon nanotubes-foam surface after ~4 m/s impact. **c**, High resolution TEM image of a typical as grown intact carbon nanotubes composing the nanofoam. **d**, TEM image of a permanently deformed CNT showing delamination and crumbling of the inner walls. **e**, TEM image of rippled and buckled nanotube. The scale bar for c-e is 5 nm.

**Figure 1**

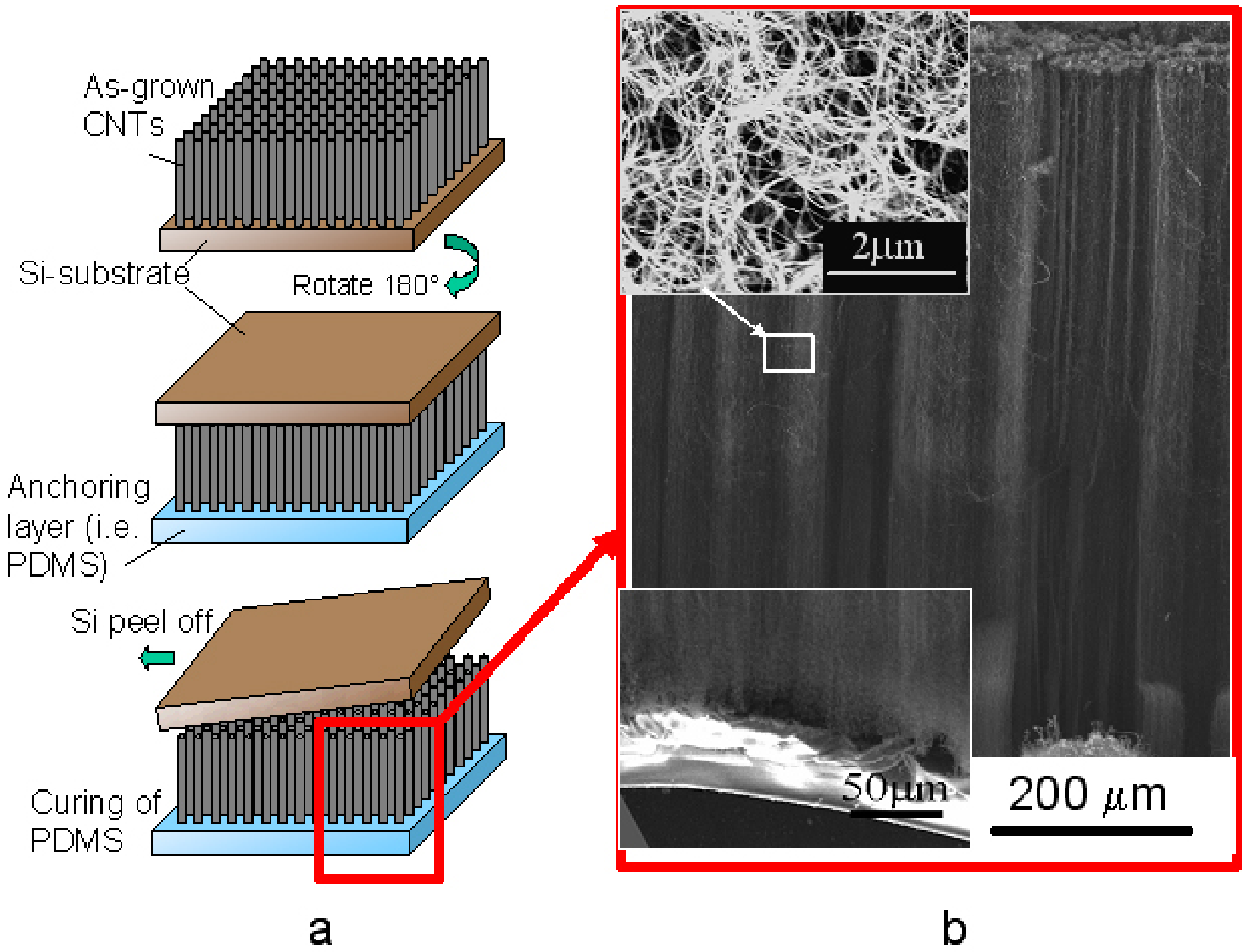
As-grown
CNTs
Si-substrate
Rotate 180°
Anchoring
layer (i.e.
PDMS)
Si peel off
Curing of
PDMS
2μm
50μm
200 μm
a
b

**Figure 2**

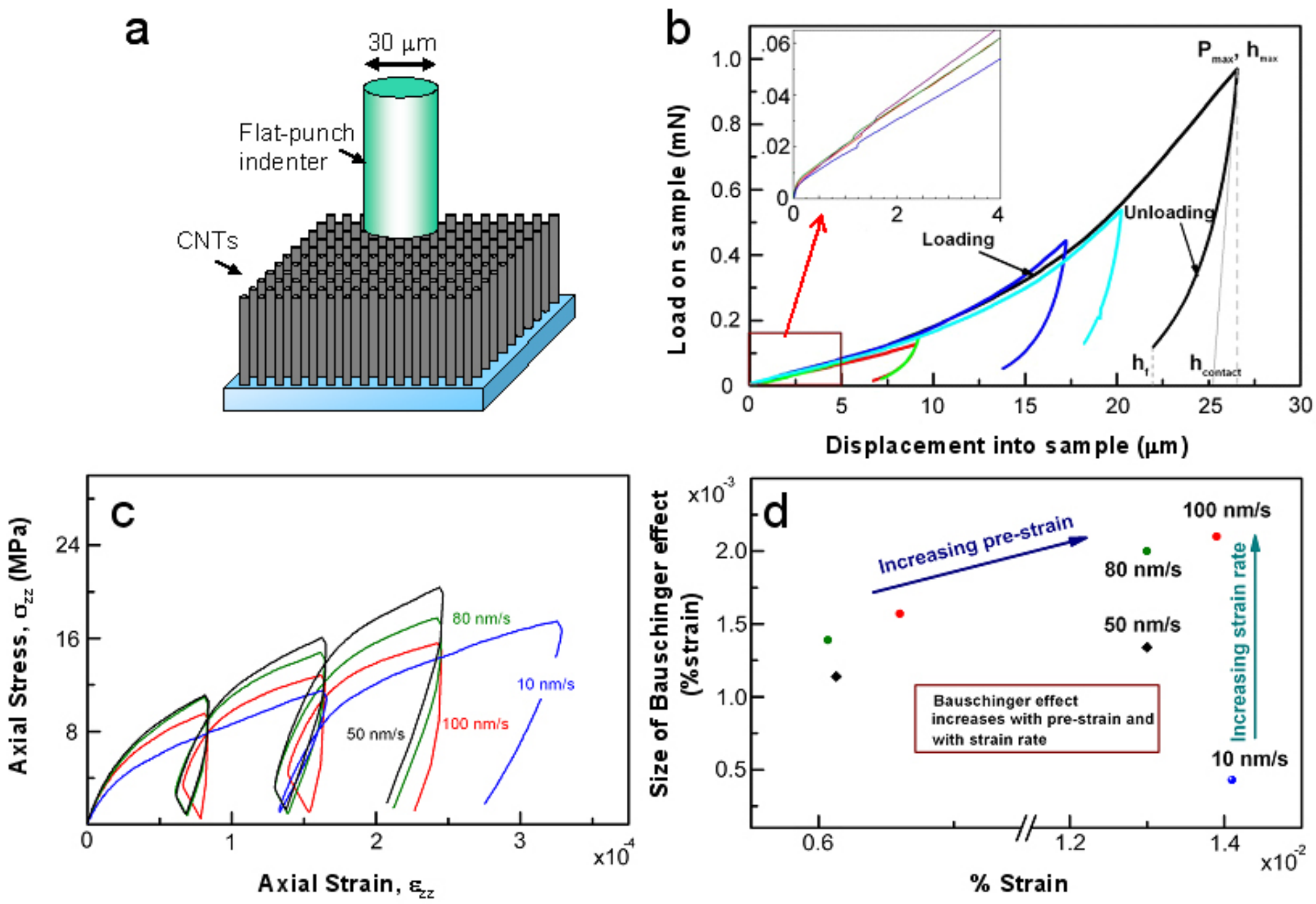
a
30 μm
Flat-punch indenter
CNTs
b
Load on sample (mN)
Displacement into sample (μm)
Pmax, hmax
Loading
Unloading
hf
hcontact
c
Axial Stress, σzz (MPa)
Axial Strain, εzz
x10-4
80 nm/s
10 nm/s
50 nm/s
100 nm/s
d
Size of Bauschinger effect (%strain)
x10-3
% Strain
x10-2
Increasing pre-strain
Increasing strain rate
100 nm/s
80 nm/s
50 nm/s
10 nm/s
Bauschinger effect increases with pre-strain and with strain rate

**Figure 3**

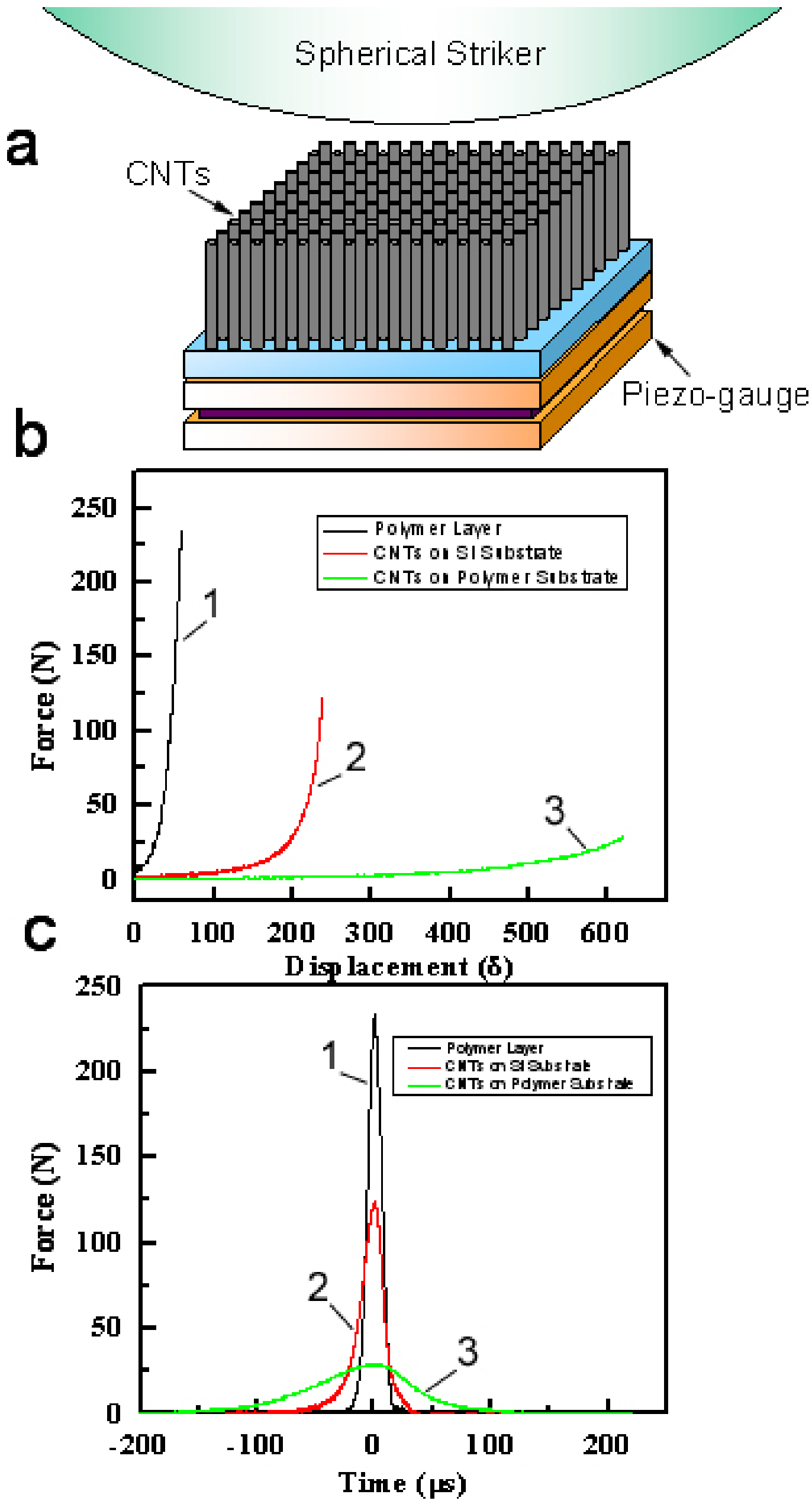
Spherical Striker
a
CNTs
Piezo-gauge
b
Polymer Layer
CNTs on SI Substrate
CNTs on Polymer Substrate
1
2
3
Force (N)
Displacement (δ)
c
Polymer Layer
CNTs on SI Substrate
CNTs on Polymer Substrate
Force (N)
Time (μs)

**Figure 4**

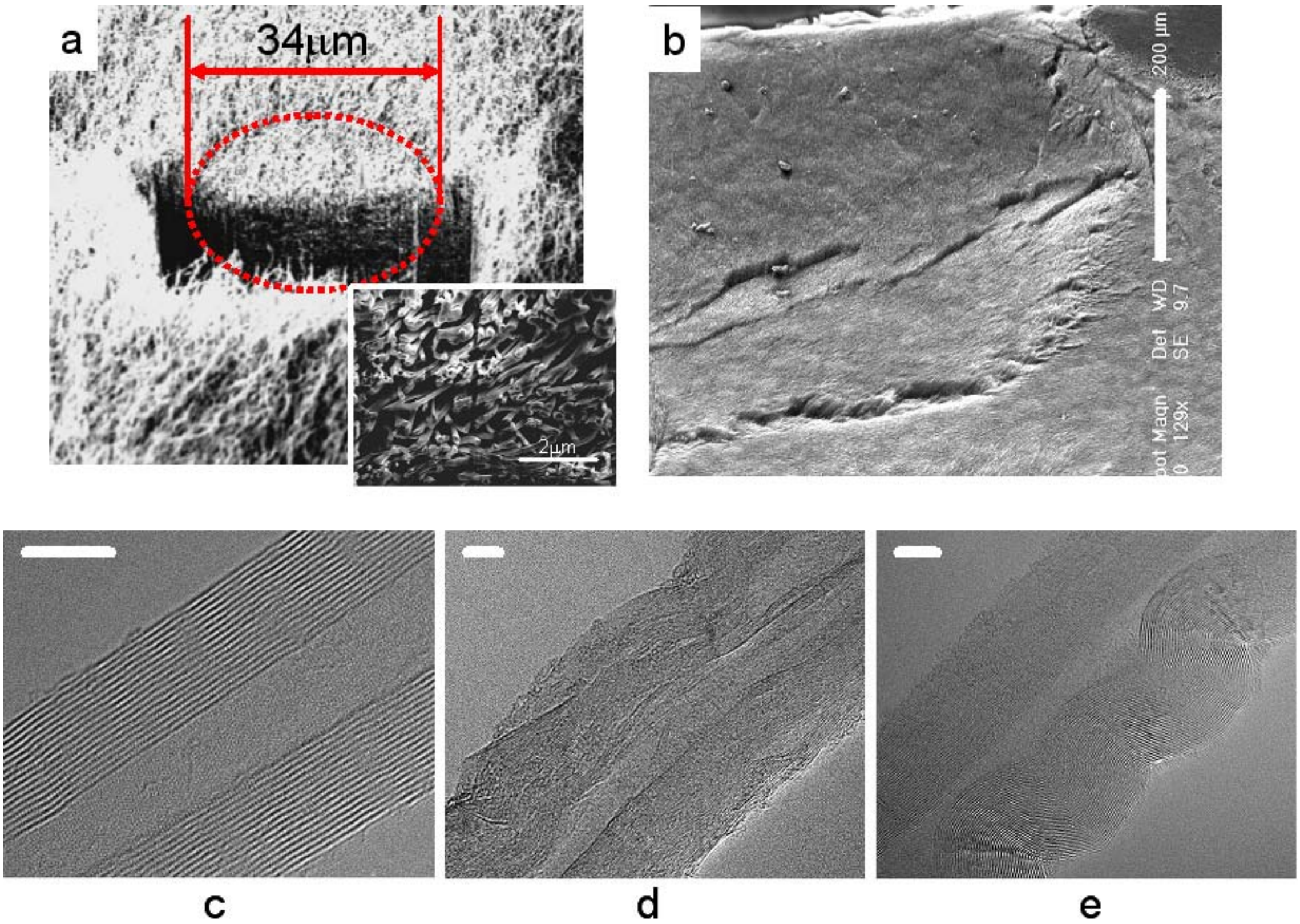